\pdfoutput=1
\documentclass[11pt]{article}

\usepackage[final]{acl}
\usepackage{times}
\usepackage{latexsym}
\usepackage[T1]{fontenc}
\usepackage[utf8]{inputenc}
\usepackage{microtype}

\usepackage{times} 
\usepackage{helvet}
\usepackage{courier} 
\usepackage{graphicx}
\usepackage{natbib} 
\usepackage{spreadtab}
\usepackage{caption}
\usepackage[utf8]{inputenc}
\usepackage[T1]{fontenc}
\usepackage[english]{babel}
\usepackage{amsfonts}
\usepackage{nicefrac}
\usepackage{microtype}
\usepackage{amsmath}
\usepackage{amssymb}
\usepackage{mathtools}
\usepackage{subcaption}
\usepackage{booktabs}
\usepackage{ragged2e}
\usepackage{tikz}
\usepackage{stackengine}
\usepackage{etoolbox}
\usepackage{xspace}
\usepackage{xpatch}
\usepackage{cuted}
\usepackage{enumerate}
\usepackage{xstring}
\usepackage{setspace}
\usepackage{tabularx}
\usepackage{makecell}
\usepackage{changepage}
\usepackage{enumitem}
\usepackage{cuted}
\usepackage{cancel}
\usepackage{eqparbox}
\usepackage{bibentry}
\usepackage[hang,flushmargin]{footmisc}
\usepackage[capitalise,noabbrev,nameinlink]{cleveref}
\usepackage[useregional=numeric]{datetime2}
\usepackage[linesnumbered,ruled,noend]{algorithm2e}
\usepackage{adjustbox}
\usepackage{threeparttable}
\usepackage{booktabs, caption, makecell}
\usepackage{colortbl}

\usepackage[switch,mathlines]{lineno}
\usepackage{lipsum}
\usepackage{etoolbox}
\usepackage{longtable}

\graphicspath{{figures/}}

\title{CITADEL: Conditional Token Interaction via Dynamic Lexical Routing\\for Efficient and Effective Multi-Vector Retrieval}

\author{Minghan Li$^1$\thanks{\ \ This work is done during Minghan's internship at Meta.}\ ,
        Sheng-Chieh Lin$^1$,
        Barlas Oguz$^2$,
        Asish Ghoshal$^2$,\\
        {\bf Jimmy Lin$^1$},
        {\bf Yashar Mehdad$^2$},
        {\bf Wen-tau Yih$^2$}, \and 
        {\bf Xilun Chen$^2$\thanks{\ \ Xilun and Minghan contributed equally to this work.}}\\[1ex]
        University of Waterloo$^1$, Meta AI$^2$\\[1ex]
        \texttt{\{m692li,s269lin,jimmylin\}@uwaterloo.ca}\\[1ex]
        \texttt{\{barlaso,aghoshal,mehdad,scottyih,xilun\}@meta.com}\\[1ex]
        }

\begin{document}
\maketitle
\begin{abstract}
Multi-vector retrieval methods combine the merits of sparse (e.g. BM25) and dense (e.g. DPR) retrievers and have achieved state-of-the-art performance on various retrieval tasks.
These methods, however, are orders of magnitude slower and need much more space to store their indices compared to their single-vector counterparts.
In this paper, we unify different multi-vector retrieval models from a token routing viewpoint and propose conditional token interaction via dynamic lexical routing, namely CITADEL, for efficient and effective multi-vector retrieval.
CITADEL learns to route different token vectors to the predicted lexical ``keys'' such that a query token vector only interacts with document token vectors routed to the same key.
This design significantly reduces the computation cost while maintaining high accuracy.
Notably, CITADEL achieves the same or slightly better performance than the previous state of the art, ColBERT-v2, on both in-domain (MS MARCO) and out-of-domain (BEIR) evaluations, while being nearly \textbf{40 times} faster.
Code and data are available at \url{https://github.com/facebookresearch/dpr-scale}.

\end{abstract}
\section{Introduction}\label{sec:intro}
\begin{figure*}[t!]
\begin{subfigure}[t]{.48\linewidth}
\centering
\includegraphics[width=1\textwidth]{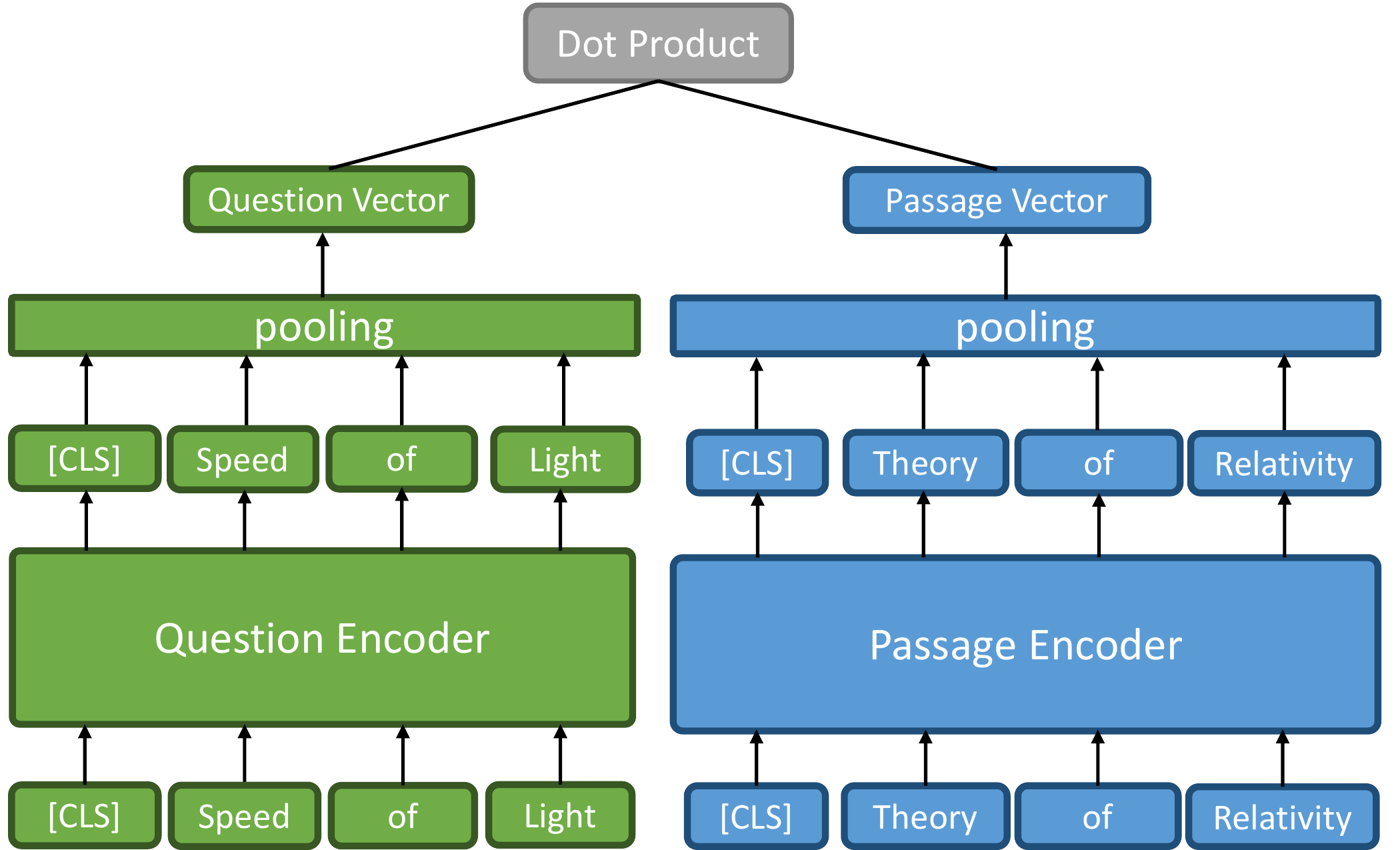}
\caption{Single-vector retriever: no routing}
\label{fig:dpr}
\end{subfigure}%
\hspace*{0.3cm} 
\begin{subfigure}[t]{.48\linewidth}
\centering
\includegraphics[width=1\textwidth]{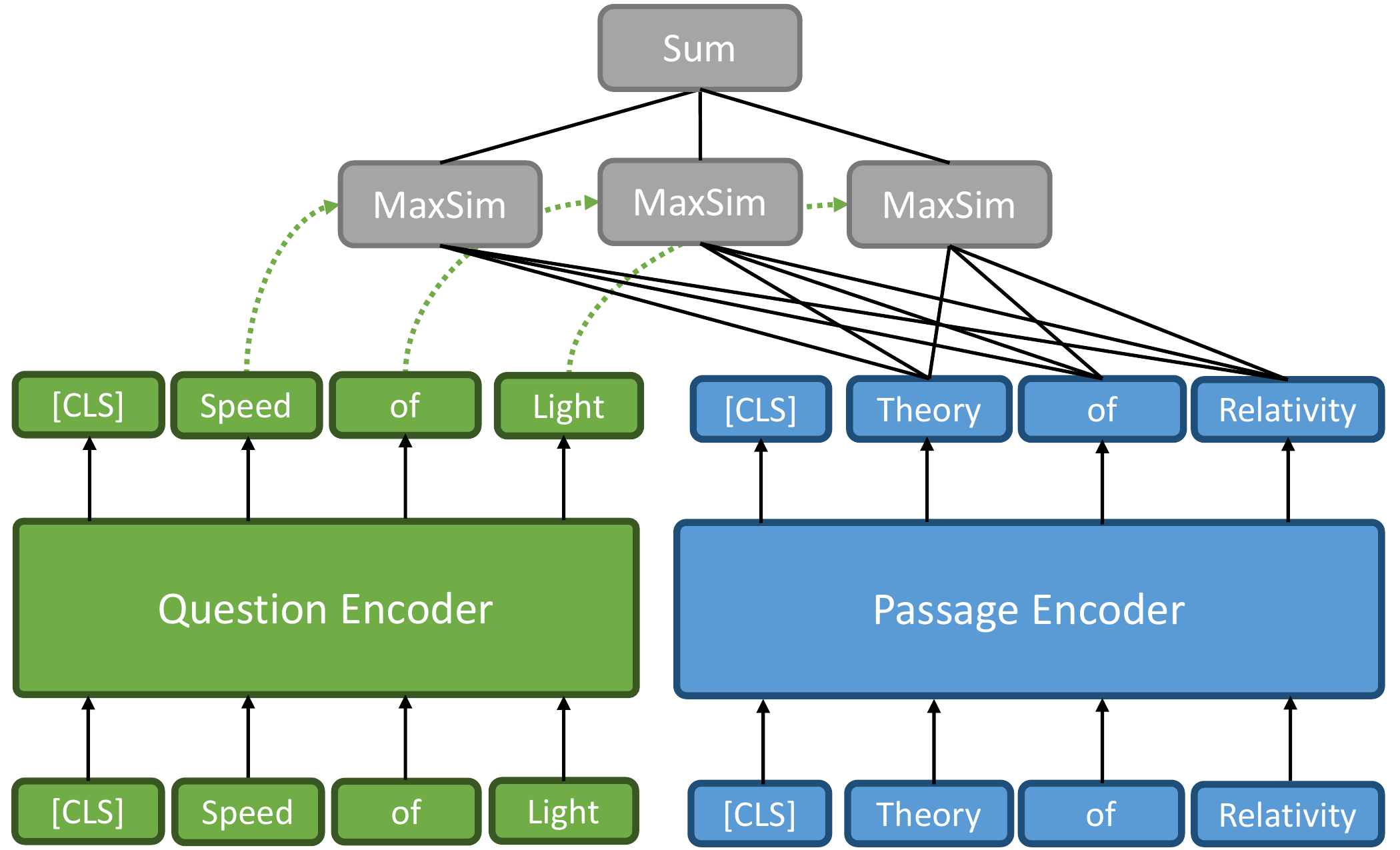}
\caption{ColBERT: all-to-all routing}
\label{fig:colbert}
\end{subfigure}\\
%
\begin{subfigure}[t]{.48\linewidth}
\centering
\includegraphics[width=1\textwidth]{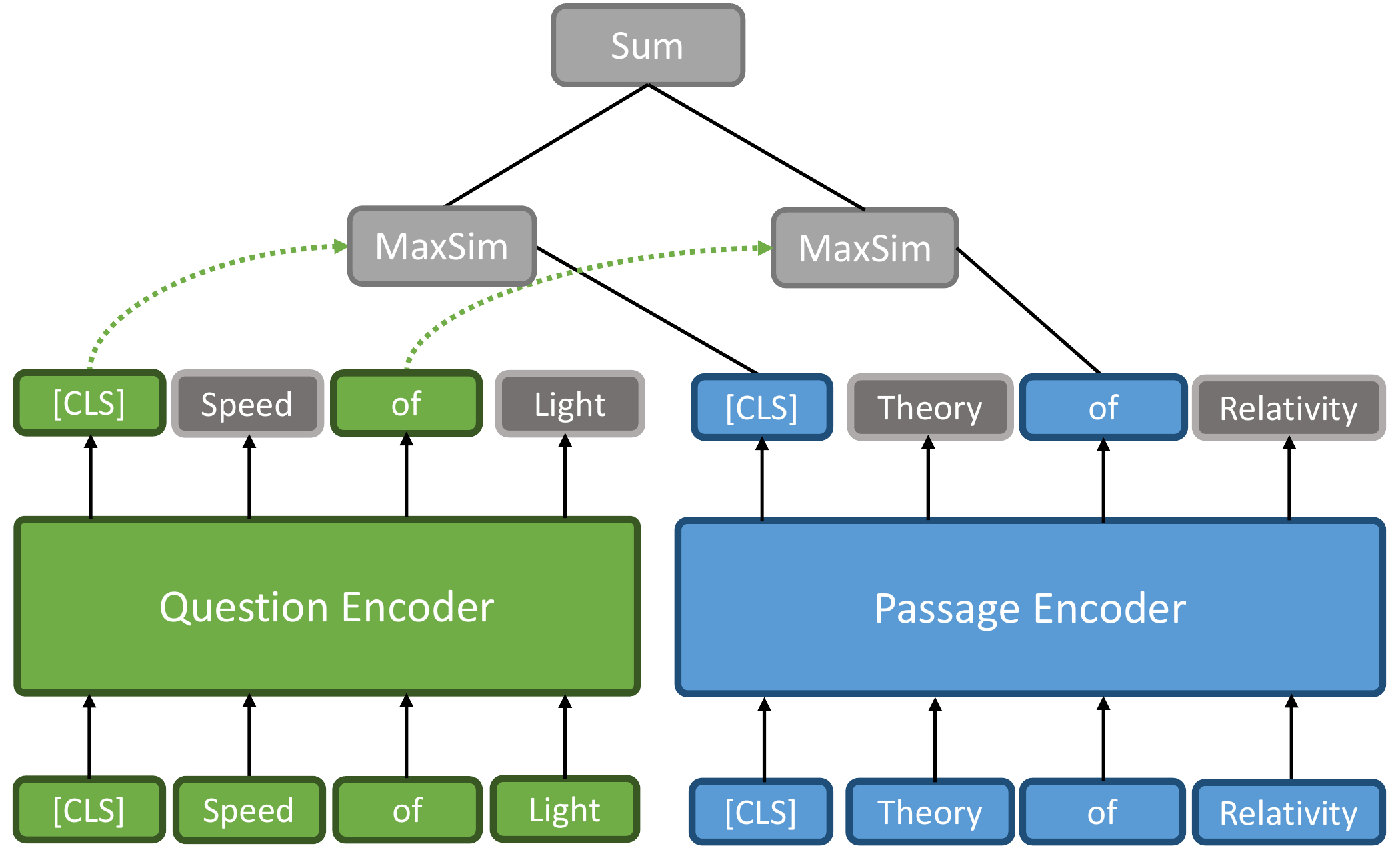}
\caption{COIL: static lexical routing using exact match}
\label{fig:coil}
\end{subfigure}%
\hspace*{0.3cm}
\begin{subfigure}[t]{.48\linewidth}
\centering
\includegraphics[width=1\textwidth]{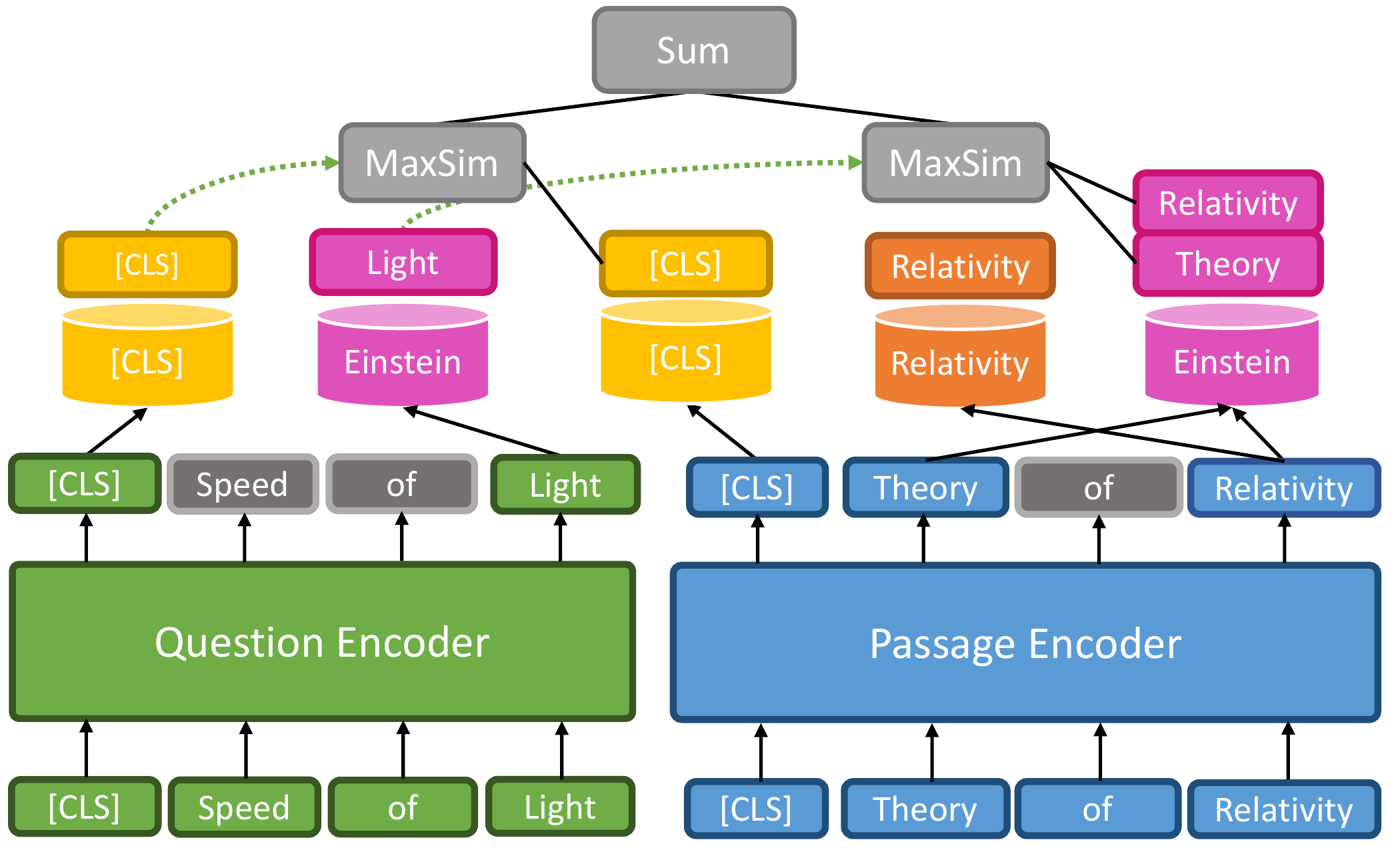}
\caption{CITADEL: dynamic lexical routing}
\label{fig:citadel}
\end{subfigure}%
\caption{A unified token routing view over different multi-vector and single-vector retrieval models.
The cylinder in (d) represents the lexical key for token routing.
The grey tokens in (c) and (d) represent the tokens that do not contribute to the final similarity.
For (d), CITADEL routes ``\textit{light}'', ``\textit{theory}'', and ``\textit{relativity}'' to the ``\textit{Einstein}'' key to avoid term mismatch using a learned routing function.
}
\label{fig:demo}
\end{figure*}

The goal of information retrieval~\citep{Manning2005IntroductionTI} is to find a set of related documents from a large data collection given a query.
Traditional bag-of-words systems~\citep{robertson2009bm25,lin2021pyserini} calculate the ranking scores based on the query terms appearing in each document, which have been widely adopted in many applications such as web search~\citep{nguyen2016ms,Brickley2019GoogleDS} and open-domain question answering~\citep{chen-etal-2017-reading,lee-etal-2019-latent}.
Recently, dense retrieval~\citep{karpukhin-etal-2020-dense} based on pre-trained language models~\citep{devlin-etal-2019-bert,Liu2019RoBERTaAR} has been shown to be very effective.
It circumvents the term mismatch problem in bag-of-words systems by encoding the queries and documents into low-dimensional embeddings and using their dot product as the similarity score (Figure~\ref{fig:dpr}).
However, dense retrieval is less robust on entity-heavy questions~\citep{sciavolino-etal-2021-simple} and out-of-domain datasets~\citep{Thakur2021BEIRAH}, therefore calling for better solutions~\citep{Formal2021SPLADESL,gao-callan-2022-unsupervised}.

\begin{figure}[t!]
\centering
\hspace*{-0.4cm} 
\includegraphics[width=.49\textwidth]{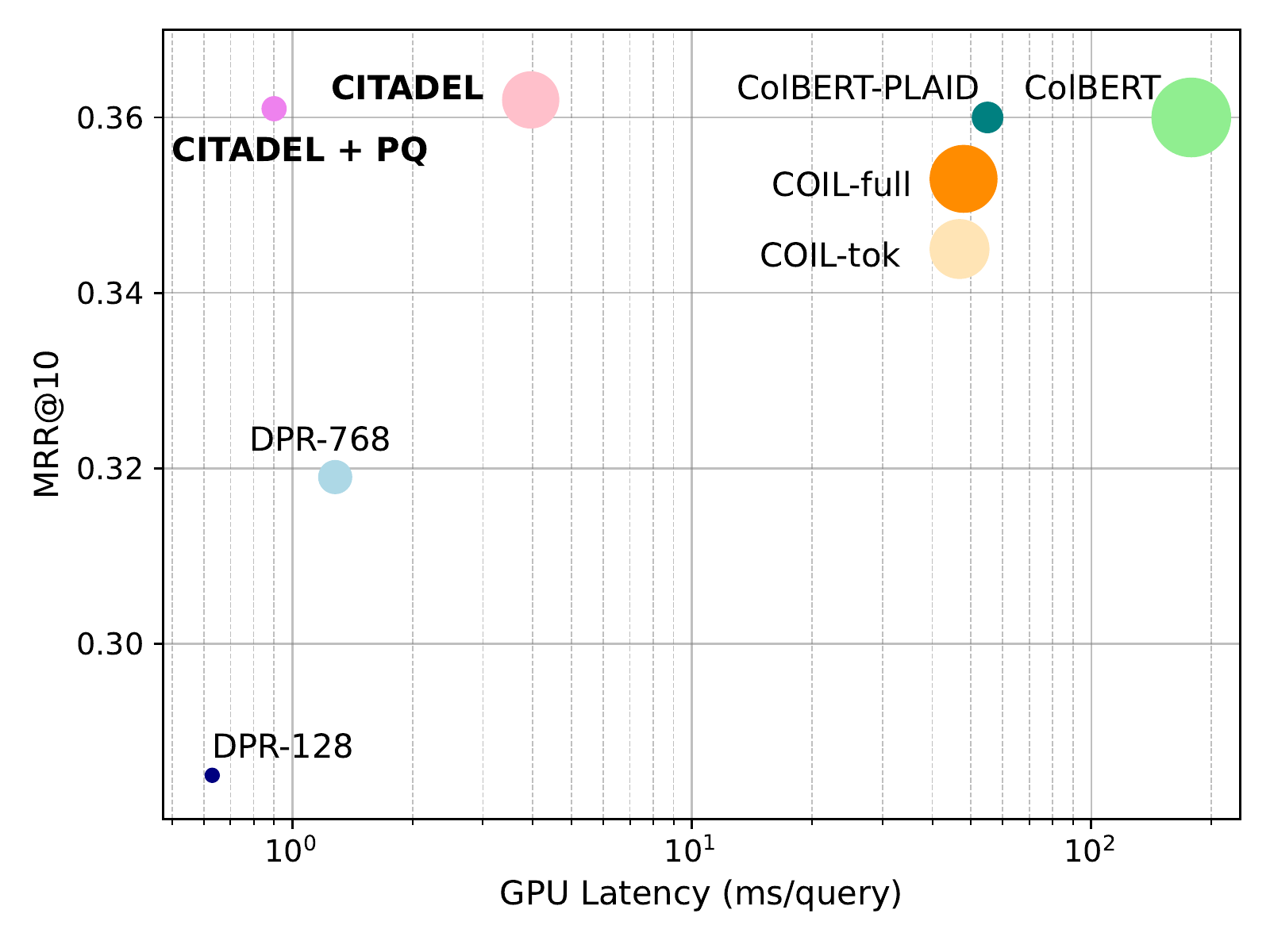}
\caption{GPU latency vs ranking performance (MRR@10) on MS MARCO passages with an A100 GPU. The \emph{circle size} represents the index storage on disk. All models are trained without hard-negative mining, distillation, or further pre-training.
}
\label{fig:showoff}
\end{figure}

In contrast, another family of retrieval models encodes input tokens into vectors first and performs fine-grained token interaction between queries and documents, known as multi-vector retrieval.
Multi-vector retrieval has shown strong performance on both in-domain and out-of-domain evaluations.
Among them, ColBERT~\citep{colbert} is arguably the most celebrated method that has been the state of the art on multiple datasets so far.
However, its wider application is hindered by its large index size and high retrieval latency.
This problem results from the redundancy in the token interaction of ColBERT where many tokens might not contribute to the sentence semantics at all.
To improve this, COIL~\citep{gao-etal-2021-coil} imposes an exact match constraint on ColBERT for conditional token interaction, where only token embeddings with the same token id could interact with each other.
Although reducing the latency, the word mismatch problem reoccurs and the model may fail to match queries and passages that use different words to express the same meaning. 

In this paper, we first give a unified view of existing multi-vector retrieval methods based on \emph{token routing} (Section~\ref{sec:background}), providing a new lens through which we expose the limitations of current models.
Under the token routing view, ColBERT could be seen as \textit{all-to-all routing}, where each query token exhaustively interacts with all passage tokens (Figure~\ref{fig:colbert}).
COIL, on the other hand, could be seen as \textit{static lexical routing} using an exact match constraint, as each query token only interacts with the passage tokens that have the same token id as the query token (Figure~\ref{fig:coil}). 
As mentioned above, this exact match heuristic improves the latency over ColBERT but negatively affects the accuracy.

In contrast, we propose a novel conditional token interaction method using \textit{dynamic lexical routing} called CITADEL as shown in Figure~\ref{fig:citadel}.
Instead of relying on static heuristics such as exact match, we train our model to dynamically moderate token interaction so that each query token only interacts with the most relevant tokens in the passage.
This is achieved by using a lexical \emph{router}, trained end-to-end with the rest of the model, to route each contextualized token embedding to a set of activated lexical ``keys'' in the vocabulary.
In this way, each query token embedding only interacts with the passage token embeddings that have the same activated key, which is dynamically determined during computation.
As we shall see in Section~\ref{sec:analysis:num_dot_product}, this learning-based routing does not lose any accuracy compared to the exhaustive all-to-all routing while using even fewer token interactions than COIL (Section~\ref{sec:method:regularization}), leading to a highly effective and efficient retriever.

Experiments on MS MARCO passages~\citep{nguyen2016ms} and TREC DeepLearning 2019/2020 show that CITADEL achieves the same level of accuracy as ColBERT-v2 and sometimes even slightly better. 
We further test CITADEL on BEIR~\citep{Thakur2021BEIRAH} for out-of-domain evaluation and CITADEL still manages to keep up with ColBERT-v2~\citep{santhanam-etal-2022-colbertv2} which is the current state of the art.
As for the latency, CITADEL can yield an average latency of 3.21 ms/query on MS MARCO passages using an A100 GPU, which is nearly 40$\times$ faster than ColBERT-v2.
By further combining with product quantization, CITADEL's index only takes 13.3 GB on MS MARCO passages and reduces the latency to 0.9 ms/query as shown in Figure~\ref{fig:showoff}.
\section{A Unified Token Routing View of Multi-Vector Retrievers}\label{sec:background}
In this section, we outline a unified view for understanding various neural retrievers, especially multi-vector ones, using the concept of token routing that dictates which query and passage tokens can interact with each other.
\subsection{Single-Vector Retrieval}\label{sec:background:dense}
Given a collection of documents and a set of queries, single-vector models~\citep{karpukhin-etal-2020-dense,contriever} use a bi-encoder structure where its query encoder $\eta_Q(\cdot)$ and document encoder $\eta_D(\cdot)$ are independent functions that map the input to a low-dimensional vector.
Specifically, the similarity score $s$ between the query $q$ and document $d$ is defined by the dot product between their encoded vectors $v_q=\eta_Q(q)$ and $v_d=\eta_D(d)$:
\begin{align}\label{eq:dpr_sim}
    s(q,d) = v_q^Tv_d.
\end{align}
As all the token embeddings are pooled (e.g., mean pooling, \texttt{[CLS]} pooling) before calculating the similarity score, no routing is committed as shown in Figure~\ref{fig:dpr}.

\subsection{Multi-Vector Retrieval}\label{sec:background:multivec}
ColBERT~\citep{colbert} proposes late interaction between the tokens in a query $q=\{q_1, q_2,\cdots,q_N\}$ and a document $d=\{d_1, d_2,\cdots,d_M\}$:
\begin{align}\label{eq:colbert_sim}
    s(q,d) =  \sum_{i=1}^N \max_{j} v_{q_i}^T v_{d_j},
\end{align}
where $v_{q_i}$ and $v_{d_j}$ denotes the last-layer contextualized token embeddings of BERT. 
This is known as the MaxSim operation where it first takes the max over the dot products between each query and document token embedding, and then sums over the score for each query token. 
As this operation needs to exhaustively compare each query token to all document tokens, we refer to this as \emph{all-to-all routing} as shown in Figure~\ref{fig:colbert}.
The latency of ColBERT is inflated by the redundancy in the all-to-all routing, as many tokens do not contribute to the sentence semantics.
This also drastically increases the storage, requiring complex engineering schemes and low-level optimization to make it more practical~\citep{santhanam-etal-2022-colbertv2,santhanam2022plaid}.

Another representative multi-vector approach known as COIL~\citep{gao-etal-2021-coil} proposes an exact match constraint on the MaxSim operation where only the embeddings with the same token id could interact with each other.
Let {\small{$\mathcal{J}_i = \{j\mid d_j = q_i, 1\leq j \leq M \}$}} be the subset of document tokens {\small{$\{d_j\}_{j=1}^M$}} that have the same token ID as query token $q_i$, then we have:
\begin{align}\label{eq:coil_tok_sim}
    s(q,d) =  \sum_{i=1}^N \max_{j\in\mathcal{J}_i} v_{q_i}^T v_{d_j},
\end{align}
It could be further combined with Equation~\eqref{eq:dpr_sim} to improve the effectiveness if there's no word overlap between the query and documents.
\begin{align}\label{eq:coil_full_sim}
    s(q, d) =  v_q^Tv_d + \sum_{i=1}^N \max_{j\in\mathcal{J}_i} v_{q_i}^T v_{d_j},
\end{align}
We refer to this token interaction as \emph{static lexical routing} as shown in Figure~\ref{fig:coil}.
As mentioned in Section~\ref{sec:intro}, the word mismatch problem could happen if $\mathcal{J}_i=\varnothing$ for all $q_i$, which affects the retrieval accuracy.
Moreover, common tokens such as ``\textit{the}'' and ``\textit{a}'' will be frequently routed, which will create much larger token indices during indexing compared to those rare words.
This bottlenecks the latency as COIL needs to frequently search overly large token indices.
\section{The CITADEL Method}\label{sec:methods}
\subsection{Dynamic Lexical Routing}\label{sec:methods:cti}
Instead of using the wasteful all-to-all routing or the inflexible heuristics-based static routing, we would like our model to dynamically select which query and passage tokens should interact with each other based on their contextualized representation, which we refer to as \emph{dynamic lexical routing}.
Formally, the routing function (or \emph{router}) routes each token to a set of lexical \textbf{keys} in the vocabulary, and is defined as $\phi:\mathbb{R}^{c}\rightarrow \mathbb{R}^{|\mathcal{V}|}$ where $c$ is the embedding dimension and $\mathcal{V}$ is the lexicon of keys.
For each contextualized token embedding, the router predicts a scalar score for each key in the lexicon indicating how relevant each token is to that key.
Given a query token embedding $v_{q_i}$ and a document token vector $v_{d_j}$, the token level router representations are $w_{q_i} = \phi(v_{q_i})$ and $w_{d_j} = \phi(v_{d_j})$, respectively.
The elements in the router representations are then sorted in descending order and truncated by selecting the top-$K$ query keys and top-$L$ document keys, which are {\small{${\{E_{q_i}^{(1)},E_{q_i}^{(2)},\cdots,E_{q_i}^{(K)}\}}$}} and {\small{${\{E_{d_j}^{(1)}, E_{d_j}^{(2)},\cdots,E_{d_j}^{(L)}\}}$}} for $q_i$ and $d_j$, respectively.
In practice, we use $K$=1 and $L$=5 as the default option which will be discussed in Section~\ref{sec:methods:ivf} and Section~\ref{sec:analysis:ablation}.
The corresponding routing weights for $q_i$ and $d_j$ are {\small{${\{w_{q_i}^{(1)},w_{q_i}^{(2)},\cdots,w_{q_i}^{(K)}\}}$}} and {\small{$\{w_{d_j}^{(1)}, w_{d_j}^{(2)},\cdots,w_{d_j}^{(L)}\}$}}, respectively.
The final similarity score is similar to Equation~\eqref{eq:coil_tok_sim}, but we substitute the static lexical routing subset $\mathcal{J}_i$ with a dynamic key set predicted by the router:
{\small{$\mathcal{E}^{(k)}_i=\{j,l\mid E_{d_j}^{(l)}=E_{q_i}^{(k)}, 1\leq j \leq M, 1\leq l \leq L\}$}}
for each key $E_{q_i}^{(k)}$ of the query token $q_i$:
\begin{align}\label{eq:citadel_sim}
s(q,d) = \sum_{i=1}^N\sum_{k=1}^K \max_{\ j, l\in\mathcal{E}^{(k)}_i} (w_{q_i}^{(k)}\cdot v_{q_i})^T (w_{d_j}^{(l)}\cdot v_{d_j}),
\end{align}
Optionally, all \texttt{[CLS]} tokens can be routed to an additional \emph{semantic} key to complement our learned lexical routing.
We then follow DPR~\citep{karpukhin-etal-2020-dense} to train the model contrastively.
Given a query $q$, a positive document $d^+$, and a set of negative documents $D^-$, the constrastive loss is:
\begin{align}
    \mathcal{L}_{\text{e}} = -\log\frac{\exp(s(q, d^+))}{\exp(s(q, d^+)) + \underset{\scalebox{0.5}{$d^-\in D^-$}}{\sum} \exp(s(q, d^-))},
\end{align}
such that the distance from the query to the positive document $d^+$ is smaller than the query to the negative document $d^-$. 

\subsection{Router Optimization}
To train the router representation $\phi(q)$ and $\phi(d)$, we adopt a contrastive loss such that the number of overlapped keys between a query and documents are large for positive $(q,d^+)$ pairs and small for negative pairs $(q,d^-)$.  
We first pool the router representation for each query and document over the tokens. 
Given a sequence of token-level router representations $\{\phi(v_1),\phi(v_2),\cdots,\phi(v_M)\}$, the sequence-level representation is defined as:
\begin{align}
    \Phi = \max_{j=1}^M \phi(v_j).
\end{align}
Similar to~\citep{Formal2021SPLADEVS}, We find max pooling works the best in practice compared to other pooling methods. 
Subsequently, the contrastive loss for training the router is:
\begin{align}\label{eq:citadel_router_optim}
    \mathcal{L}_\text{r} = -\log\frac{\exp(\Phi_q^T\Phi_{d^+})}{\exp(\Phi_q^T\Phi_{d^+}) + \underset{\scalebox{0.5}{$d^-\in D^-$}}{\sum}\exp(\Phi_q^T\Phi_{d^-})},
\end{align}

\subsection{Sparsely Activated Router Design}
Softmax activation is commonly used for computing the routing weights in conditional computation models~\citep{Fedus2021SwitchTS,Mustafa2022MultimodalCL}.
However, softmax often yields a small probability over a large number of dimensions (in our case, about 30,000) and the product of two probability values are even smaller, which makes it not suitable for yielding the routing weights $w_{q_i}^{(k)}$ and $w_{d_j}^{(l)}$ in Equation~\eqref{eq:citadel_sim} as the corresponding gradients are too small.
Instead, we use the activation from SPLADE~\citep{Formal2021SPLADESL,Formal2021SPLADEVS} to compute the router representation for a token embedding $v_j$:
\begin{align}\label{eq:citadel_router}
    \phi(v_j) = \log(1+\text{ReLU}(W^Tv_j+b)),
\end{align}
where $W$ and $b$ are the weights and biases of the Masked Language Modeling (MLM) layer of BERT.
The SPLADE activation brings extra advantages as the $\text{ReLU}$ activation filters irrelevant keys while the log-saturation suppresses overly large ``wacky'' weights~\citep{mackenzie2021wacky}.

\subsection{Regularization for Routing}\label{sec:method:regularization}
\paragraph{$\ell_1$ Regularization.} Routing each token to more than one key increases the overall size of the index.
Therefore, we propose to use $\ell_1$ regularization on the router representation to encourage the router to only keep the most meaningful token interaction by pushing more routing weights to 0:
\begin{align}
    \mathcal{L}_{\text{s}} = \frac{1}{B}\sum^B_{i=1}\sum_{j=1}^T\sum_{k=1}^{|\mathcal{V}|} \phi(v_{ij})^{(k)},
\end{align}
where $|\mathcal{V}|$ is the number of keys, $B$ is the batch size, and $T$ is the sequence length.
As shown in Figure~\ref{fig:facebook_example}, CITADEL has a sparsely activated set of keys, by routing important words to multiple lexical keys while ignoring many less salient words, leading to effective yet efficient retrieval.

\paragraph{Load Balancing.} 
As mentioned in Section~\ref{sec:background:multivec}, the retrieval latency of COIL is bottlenecked by frequently searching overly large token indexes.
This results from the static lexical routing where common ``keys'' have a larger chance to be activated, which results in large token indices during indexing.
Therefore, a vital point for reducing the latency of multi-vector models is to evenly distribute each token embedding to different keys.
Inspired by Switch Transformers~\cite{Fedus2021SwitchTS}, we propose to minimize the load balancing loss that approximates the expected ``evenness'' of the number of tokens being routed to each key:
\begin{align}\label{eq:load_balance}
    \mathcal{L}_{\text{b}} = \sum_{k=1}^{|\mathcal{V}|} f_k \cdot p_k, 
\end{align}
$p_k$ is the batch approximation of the marginal probability of a token embedding being routed to the $k$-th key:
\begin{align}
    p_k = \frac{1}{B}\sum_{i=1}^B\sum_{j=1}^T \frac{\exp(W_k^Tv_{ij}+b_k)}{\sum_{k'}\exp(W_{k'}^Tv_{ij}+b_{k'})},
\end{align}
where $W$ and $b$ are the weights and bias of the routing function in Equation~\eqref{eq:citadel_router} and $v_{ij}$ is the $j$-th token embedding in sample $i$ of the batch.
$f_k$ is the batch approximation of the total number of tokens being dispatched to the $k$-th key:
\begin{align}
    f_k = \frac{1}{B}\sum_{i=1}^B\sum_{j=1}^T \mathbb{I}\{\text{argmax}(p_{ij}) = k\},
\end{align}
where $p_{ij}=\text{softmax}(W^Tv_{ij}+b)$.
Finally, we obtain the loss for training CITADEL:
\begin{align}\label{eq:loss}
    \mathcal{L} = \mathcal{L}_{\text{e}} + \mathcal{L}_{\text{r}} + \alpha \cdot \mathcal{L}_{\text{b}} + \beta \cdot \mathcal{L}_{\text{s}},
\end{align}
where $\alpha \geq 0$ and $\beta \geq 0$. The $\ell_1$ and load balancing regularization is applied on both queries and documents.

\begin{table*}[t]
\centering
\begin{adjustbox}{max width=\textwidth}
\begin{threeparttable}[t]
\begin{tabular}{l|cc|cc|cc|cc|ccc}
\toprule
\textbf{Models}& \multicolumn{2}{c|}{\textbf{MARCO Dev}}& \multicolumn{2}{c|}{\textbf{TREC DL19}}& \multicolumn{2}{c|}{\textbf{TREC DL20}}  & \multicolumn{2}{c|}{\textbf{Index Storage}}& \multicolumn{3}{c}{\textbf{Latency (ms/query)}}\\
& \small{MRR@10} &\small{R@1k} &\small{nDCG@10} &\small{R@1k}&\small{nDCG@10} &\small{R@1k} &\small{Disk (GB)} &\small{Factor$^1$} &\small{Encode (GPU)} &\small{Search (GPU)} &\small{Search (CPU)}\\
\midrule
\multicolumn{11}{c}{\qquad \textit{Models trained with only BM25 hard negatives}}\\\midrule
BM25 & \cellcolor{red!7}0.188 &\cellcolor{red!7}0.858 &\cellcolor{red!7}0.506 &\cellcolor{red!7}0.739 &\cellcolor{red!7}0.488 &\cellcolor{red!7}0.733 &\underline{\textbf{0.67}} & $\times$\underline{\textbf{0.22}} &$\times$ &$\times$ & \underline{\textbf{40.1}}\\
DPR-128 &\cellcolor{red!7}0.285 &\cellcolor{red!7}0.937 &\cellcolor{red!7} 0.576 &\cellcolor{red!7}0.702 &\cellcolor{red!7}0.603 &\cellcolor{red!7}0.757 &4.33 &$\times$1.42 & 7.09 &\underline{\textbf{0.63}} &430\\
DPR-768 &\cellcolor{red!7}0.319 &\cellcolor{red!7}0.941 &\cellcolor{red!7}0.611 &\cellcolor{red!7}0.742 &\cellcolor{red!7}0.591 &\cellcolor{red!7}0.796 &26.0 &$\times$8.52 & 7.01 &1.28 &2015\\
SPLADE &\cellcolor{red!7}0.340 &\cellcolor{red!7}0.965 &\cellcolor{green!7}0.683 &\cellcolor{green!7}0.813 &\cellcolor{green!7}0.671 &\cellcolor{green!7}0.823 &2.60 & $\times$0.85 &7.13 & $\times$ & 475\\
COIL-tok &\cellcolor{red!7}0.350 &\cellcolor{red!7}0.964 &\cellcolor{green!7}0.660 &\cellcolor{green!7}0.809 &\cellcolor{green!7}0.679 &\cellcolor{green!7}0.825 &52.5 & $\times$17.2 &10.7 & 46.8 &1295\\
COIL-full &\cellcolor{red!7}0.353 &\cellcolor{green!7}0.967 &\cellcolor{green!7}\underline{0.704} &\cellcolor{green!7}\underline{0.835} &\cellcolor{green!7}\underline{0.688} &\cellcolor{green!7}\underline{0.841} &78.5 & $\times$25.7 &10.8 &47.9 &3258\\
ColBERT &\cellcolor{green!7}0.360 &\cellcolor{green!7}0.968 &\cellcolor{green!7}0.694 &\cellcolor{green!7}0.830 &\cellcolor{green!7}0.676 &\cellcolor{green!7}0.837  &154 &$\times$50.5 &10.9  & 178 &-  \\
CITADEL&\cellcolor{green!7}\underline{0.362} &\cellcolor{green!7}\underline{0.975} &\cellcolor{green!7}0.687 &\cellcolor{green!7}0.829 &\cellcolor{green!7}0.661 &\cellcolor{green!7}0.830  &78.3 &$\times$25.7 &10.8 &3.95 & 520\\
\midrule
\multicolumn{11}{c}{\qquad \textit{Models trained with further pre-training/hard-negative mining/distillation}}\\\midrule
coCondenser &\cellcolor{red!7}0.382 &\cellcolor{green!7}0.984 &\cellcolor{green!7}0.674 &\cellcolor{green!7}0.820  &\cellcolor{green!7}0.684 &\cellcolor{green!7}0.839 &26.0 &$\times$8.52 & 7.01 &\underline{1.28} &2015\\
SPLADE-v2 & \cellcolor{red!7}0.368 &\cellcolor{green!7}0.979 &\cellcolor{green!7}0.729 &\cellcolor{green!7}0.865 &\cellcolor{green!7}0.718 &\cellcolor{green!7}0.890 &4.12 &$\times$1.35 &7.13 &$\times$&2710\\
ColBERT-v2 &\cellcolor{green!7}0.397 &\cellcolor{green!7}\underline{\textbf{0.985}} &\cellcolor{green!7}\underline{\textbf{0.744}} &\cellcolor{green!7}\underline{\textbf{0.882}} &\cellcolor{green!7}\underline{\textbf{0.750}} &\cellcolor{green!7}\underline{\textbf{0.894}} &29.0 &$\times$9.51 &10.9 &122 &3275\\
ColBERT-PLAID$^2$ &\cellcolor{green!7}0.397 &\cellcolor{green!7}0.984 &\cellcolor{green!7}0.744 &\cellcolor{green!7}0.882  &\cellcolor{green!7}0.749 &\cellcolor{green!7}0.894 &\underline{22.1}  &$\times$\underline{7.25} &10.9 &55.0 &\underline{370} \\
CITADEL$^+$ &\cellcolor{green!7}{\underline{\textbf{0.399}}} &\cellcolor{green!7}0.981 &\cellcolor{green!7}0.703 &\cellcolor{green!7}0.830  &\cellcolor{green!7}0.702 &\cellcolor{green!7}0.859 &81.3  &$\times$26.7 &10.8 &3.21 &{635} \\
\bottomrule
\end{tabular}
\begin{tablenotes}\footnotesize
\item[1] Factor: Ratio of index size to plain text size.
\item[2] The PLAID implementation of ColBERT contains complex engineering schemes and low-level optimization such as centroid interaction and fast kernels.
\end{tablenotes}
\end{threeparttable}
\end{adjustbox}
\caption{In-domain evaluation on MS MARCO passages and TREC DL 2019/2020. CITADEL$^+$ is trained with cross-encoder distillation and hard-negative mining. The red region means CITADEL/CITADEL$^+$ is better than the method while the green region means that there's no statistical significance ($p > 0.05$). ``$\times$'' means not implemented and ``-'' means not practical to evaluate on a single CPU.
}
\label{tbl:msmarco}
\end{table*}
\subsection{Inverted Index Retrieval}\label{sec:methods:ivf}
As we store the token embeddings according to their lexical keys, this is essentially inverted index retrieval like BM25 but instead of a scalar for each token, we use a low-dimensional vector and the doc product as the term relevance.
\paragraph{Indexing and Post-hoc Pruning.} To further reduce index storage, we can prune the vectors with routing weights less than a certain threshold $\tau$ after training.
For a key $E$ in the lexicon $\mathcal{V}$, the token index $\mathcal{I}_E$ consists of token embeddings $v_{d_j}$ and the routing weight $w^{E}_{d_j}$ for all documents $d$ in the corpus $\mathcal{C}$ is:
\begin{align}\label{eq:pruning}
    \mathcal{I}_{E} = \{w^{E}_{d_j} \cdot v_{d_j} \mid &w^{E}_{d_j} > \tau,1\leq j\leq M, \forall d\in\mathcal{C} \}.
\end{align}
We will discuss the impact of post-hoc pruning in Section~\ref{sec:analysis:pruning}, where we find that post-hoc pruning can reduce the index size by 3$\times$ without significant accuracy loss.
The final search index is defined as $\mathcal{I} = \{\mathcal{I}_{E}| E\in \mathcal{V}\}$, where the load-balancing loss in Equation~\eqref{eq:load_balance} will encourage the size distribution over $\mathcal{I}_{E}$ to be as even as possible. 
In practice, we set the number of maximal routing keys of each token to 5 for the document and 1 for the query.
The intuition is that the documents usually contain more information and need more key capacity, which is detailedly discussed in Section~\ref{sec:analysis:ablation}.
Moreover, we find that using more than 1 routing key for the query does not improve the accuracy but increases the latency rapidly.

\paragraph{Token Retrieval.} Given a query $q$, CITADEL first encodes it into a sequence of token vectors {{$\{v_{q_i}\}_{i=1}^N$}}, and then route each vector to its top-1 key $E$ with a routing weight {{$w_{q_i}^{E}$}}.
The final representation {{$w_{q_i}^E\cdot v_{q_i}$}} will be sent to the corresponding token index $\mathcal{I}_E$ for vector search.
The final ranking list will be merged from each query token's document ranking according to Equation~\eqref{eq:citadel_sim}.
\begin{table*}[t]
\centering
\begin{adjustbox}{max width=\textwidth}
    \centering
    \begin{tabular}{l|ccccccccccccc|c}
    \toprule
        Methods & AA & CF & DB & Fe & FQ & HQ & NF & NQ & Qu & SF & SD & TC & T2 & Avg. \\\midrule
\multicolumn{15}{c}{\qquad \textit{Models trained with only BM25 hard negatives}}\\\midrule
        BM25 & 0.315 & 0.213 & 0.313 & 0.753 & 0.236 & 0.603 & 0.325 & 0.329 & 0.789 & 0.665 & 0.158 & 0.656 & \underline{\textbf{0.367}} & 0.440 \\ 
        DPR-768 & 0.323 & 0.167 & 0.295 & 0.651 & 0.224 & 0.441 & 0.244 & 0.410 & 0.75 & 0.479 & 0.103 & 0.604 & 0.185 & 0.375 \\
        SPLADE &0.445 &0.201 &0.370 &0.740 &0.289 &0.640 &0.322 &0.469 &0.834 &0.633 &0.149 &0.661 &0.201 &0.453\\
        COIL-full & 0.295 & \underline{0.216} & 0.398 & \underline{\textbf{0.840}} &0.313 & \underline{\textbf{0.713}} &\underline{0.331} & 0.519 & 0.838 & \underline{\textbf{0.707}} & 0.155 & 0.668 & 0.281 & 0.483 \\
        ColBERT &0.233 & 0.184 & 0.392 &0.771 &\underline{0.317} &0.593 &0.305 &\underline{0.524} &\underline{0.854} &0.671 &\underline{0.165} & 0.677 &0.202 &0.453 \\
        CITADEL &\underline{0.503} &0.191 &\underline{0.406} &0.784 &0.298 &0.653 &0.324 &0.510 &0.844 &0.674 &0.152  &\underline{0.687} &0.294 &\underline{0.486}\\
        \midrule
\multicolumn{15}{c}{\qquad \textit{Models with further pre-training/hard-negative mining/distillation}}\\\midrule
        coCondenser &0.440 &0.133 &0.347 &0.511 &0.281 &0.533 &0.319 &0.467 &0.863 &0.591 &0.130 &0.708 &0.143 &0.420 \\
        SPLADE-v2 & 0.479 & \underline{\textbf{0.235}} & 0.435 & \underline{{0.786}} & 0.336 & \underline{{0.684}} & 0.334 & 0.521 & 0.838 & 0.693 & 0.158 & 0.710 & 0.272 & 0.499 \\ 
        ColBERT-v2 & 0.463 & 0.176 & \underline{\textbf{0.446}} & 0.785 & \underline{\textbf{0.356}} & 0.667 & \underline{\textbf{0.338}} & \underline{\textbf{0.562}} & \underline{\textbf{0.854}} & 0.693 & \underline{\textbf{0.165}} & \underline{\textbf{0.738}} & 0.263 & 0.500 \\ 
        CITADEL$^+$ &0.490 &0.181 &0.420 &0.747 &0.332 &0.652 &0.337 &0.539 &0.852 &\underline{{0.695}} &0.147 &0.680 &0.340 &0.493 \\
        CITADEL$^+$ {\small{(w/o reg.)}} & \underline{\textbf{0.511}} & 0.182 & 0.422 & 0.765 & 0.330 & 0.664 & 0.337 & 0.540 & 0.853 & 0.690 & 0.159 & 0.715 & \underline{{0.342}} &\underline{\textbf{0.501}} \\ 
        \bottomrule
    \end{tabular}
\end{adjustbox}
\caption{Out-of-domain evaluation on BEIR benchmark. nDCG@10 score is reported. Dataset Legend~\citep{chen2022spar}: TC=TREC-COVID, NF=NFCorpus, NQ=NaturalQuestions, HQ=HotpotQA, FQ=FiQA, AA=ArguAna, T2=Touché-2020 (v2), Qu=Quora, DB=DBPedia, SD=SCIDOCS, Fe=FEVER, CF=Climate-FEVER, SF=SciFact.}
\label{tbl:beir}
\end{table*}
\section{Experiments}\label{sec:exp}
\subsection{MS MARCO Passages Retrieval}
We evaluate MS MARCO passages~\citep{nguyen2016ms} and its shared tasks, TREC DeepLearning (DL) 2019/2020 passage ranking tasks~\citep{craswell2020overview}.
The MS MARCO passages corpus has around 8.8 million passages with an average length of 60 words.
TREC DL 2019 and 2020 contain 43 and 54 test queries whose relevance sets are densely labelled with scores from 0 to 4.
Following ColBERT-v2~\citep{santhanam-etal-2022-colbertv2} and SPLADE-v2~\citep{Formal2021SPLADEVS}, we train CITADEL and other baseline models on MS MARCO passages and report the results on its dev-small set and TREC DL 2019/2020 test queries.
The evaluation metrics are MRR@10, nDCG@10, and Recall@1000 (i.e., R@1K).
We provide a detailed implementation of CITADEL and other baselines in Appendix~\ref{apx:baselines},~\ref{apx:training}, and~\ref{apx:inference}.

Table~\ref{tbl:msmarco} shows the in-domain evaluation results on MS MARCO passage and TREC DL 2019/2020. 
We divide the models into two classes: ones trained with only labels and BM25 hard negatives and the others trained with further pre-training~\citep{gao-callan-2022-unsupervised}, hard negative mining~\citep{ance2021xiong}, or distillation from a cross-encoder\footnote{\url{https://huggingface.co/cross-encoder/ms-marco-MiniLM-L-6-v2}}.
CITADEL is trained only with BM25 hard negatives, while CITADEL$^+$ is trained with cross-encoder distillation and one-round hard negative mining.
The default pruning threshold is $\tau=0.9$.
As shown in Section~\ref{sec:analysis:pruning}, $\tau$ can be adjusted to strike different balances between latency, index size and accuracy.
In both categories, CITADEL/CITADEL$^+$ outperforms the baseline models on the MS MARCO passages dev set and greatly reduces the search latency on both GPU and CPU.
For example, CITADEL$^+$ achieves an average latency of 3.21 ms/query which is close to DPR-768 (1.28 ms/query) on GPU, while having a 25\%  higher MRR@10 score.
CITADEL also maintains acceptable index sizes on disk, which can be further reduced using product quantization (Section~\ref{sec:analysis:ablation}).
Although not able to outperform several baselines on TREC DL 2019/2020, we perform t-test (p < 0.05) on CITADEL and CITADEL$^+$ against other state-of-the-art baselines in their sub-categories and show there is no statistical significance. 
The inconsistency is probably due to the small test set of TREC DL as it is very expensive to label hundreds of passages paired with each query. 

\subsection{BEIR: Out-of-Domain Evaluation}
We evaluate on BEIR benchmark~\citep{Thakur2021BEIRAH} which consists of a diverse set of 18 retrieval tasks across 9 domains.  
Following previous works~\citep{santhanam-etal-2022-colbertv2,Formal2021SPLADEVS}, we only evaluate 13 datasets for license reasons.
Table~\ref{tbl:beir} shows the zero-shot evaluation results on BEIR.
Without any pre-training or distillation, CITADEL manages to outperform all baselines in their sub-categories in terms of the average score.
Compared with the distilled/pre-trained models, CITADEL$^+$ still manages to achieve comparable performance.
Interestingly, we find that if no regularization like load balancing and L1 is applied during training, CITADEL$^+$ can reach a much higher average score that even outperforms ColBERT-v2.
Our conjecture is that the regularization reduces the number of token interactions and the importance of such token interaction is learned from training data. 
It is hence not surprising that the more aggressively we prune token interaction, the more likely that it would hurt out-of-domain accuracy that's not covered by the training data.

\begin{table}[!t]
\centering
\begin{adjustbox}{max width=0.49\textwidth}
\begin{tabular}{l|cc}
\toprule
Models & MRR@10 & \#DP $\times 10^6$\\ 
\midrule
ColBERT &0.360 &4213\\
COIL-full &0.353 &45.6\\
CITADEL &0.362 &10.5\\
DPR-128 &0.285 &8.84 \\
\bottomrule
\end{tabular}
\end{adjustbox}
\caption{Maximal number of dot products per query on MS MARCO passages. \# DP: number of dot products.}
\label{tbl:token_interact}
\end{table}
\section{Performance Analysis}\label{sec:analysis}
\subsection{Number of Token Interactions}\label{sec:analysis:num_dot_product}
The actual latency is often impacted by engineering details and therefore FLOPS is often considered for comparing efficiency agnostic to the actual implementation. 
In our case, however, FLOPS is impacted by the vector dimension in the nearest neighbour search which is different across models. 
Therefore, we only compare the maximal number of dot products needed as a proxy for token interaction per query during retrieval as shown in Table~\ref{tbl:token_interact}.
The number of dot products per query in CITADEL with pruning threshold $\tau=0.9$ is comparable to DPR-128 and much lower than ColBERT and COIL, which is consistent with the latency numbers in Table~\ref{tbl:msmarco}.
The reason is that CITADEL has a balanced and small inverted index credited to the $\ell_1$ regularization and the load balancing loss. 

\begin{figure}[t!]
\centering
\vspace{-0.2cm}
\hspace{-0.4cm}
\includegraphics[width=0.5\textwidth]{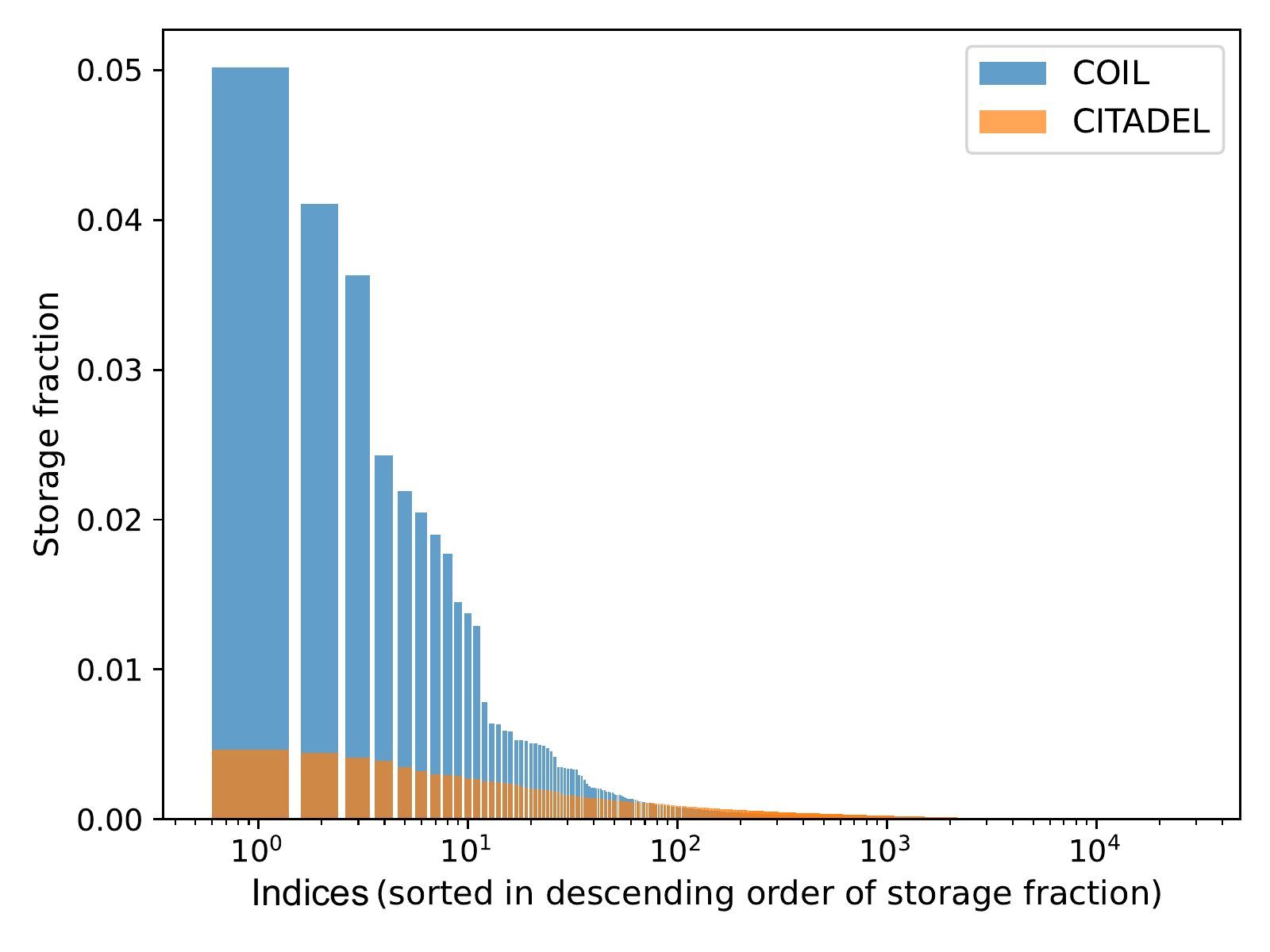}
\caption{Normalized disk storage over token indices in an inverted list of CITADEL and COIL.
}
\label{fig:index_size}
\end{figure}
\begin{figure}[t!]
\centering
\vspace{-0.3cm}
\hspace{-0.4cm} 
\includegraphics[width=.5\textwidth]{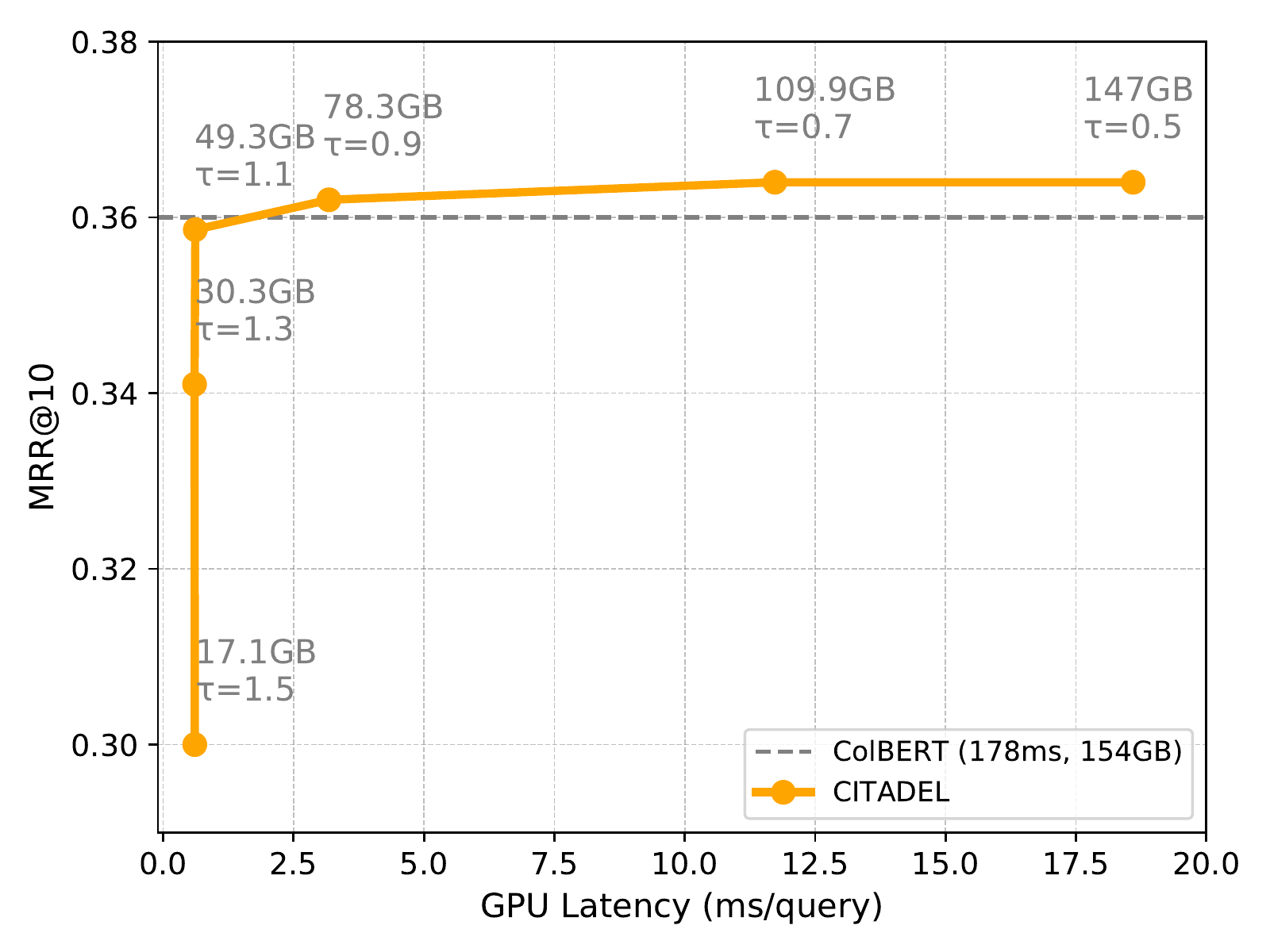}
\caption{Latency-memory-accuracy tradeoff on MS MARCO passages using post-hoc pruning. $\tau$ is the pruning threshold.
}
\label{fig:latency_tradeoff}
\end{figure}
\subsection{Effects of Load Balancing}
Figure~\ref{fig:index_size} shows why a balanced inverted index greatly speeds up the inverted index search. 
As mentioned in Section~\ref{sec:methods:ivf}, both CITADEL and COIL organize their token embeddings in an inverted list, except that COIL uses the input id of the token embedding as the key while CITADEL uses the router predictions.
This makes COIL's index distribution predetermined by the token distribution of the corpus, where frequent words like ``the'' and ``a'' will have much larger index sizes than those rare words.
In contrast, by applying the load balancing loss on the router prediction, CITADEL yields a more balanced and even index distribution where its largest index is 8 $\times$ smaller than COIL's after normalization.
We also provide a detailed latency breakdown in~\ref{apx:inference}.

\subsection{Latency-Memory-Accuracy Trade-Off}\label{sec:analysis:pruning}
\paragraph{Effects of Post-hoc Pruning.} Figure~\ref{fig:latency_tradeoff} shows the tradeoff among latency, memory, and MRR@10 on MS MARCO passages with post-hoc pruning.
We try the pruning thresholds [0.5, 0.7, 0.9, 1.1, 1.3, 1.5].
We could see that the MRR@10 score barely decreases when we increase the threshold to from 0.5 to 1.1, but the latency decreases by a large margin, from about 18 ms/query to 0.61 ms/query.
The sweet spot ($\tau=1.1$) is around (0.359 MRR@10, 49.3GB, 0.61 ms/query) which has a similar latency as DPR-128 while being as effective as ColBERT.
Another sweet spot ($\tau=0.9$) is around (0.362, 78.5GB, 3.95 ms/query), which has an even better MRR@10 score than ColBERT and only increases the storage and latency by a small gap. 
This simple pruning strategy is extremely effective and we shall see later in Section~\ref{sec:interpret} it also yields interpretable document representations.
\paragraph{Combination with Product Quantization. } We could further reduce the latency and storage with product quantization~\citep{jegou2011pq} (PQ) as shown in Table~\ref{tbl:pq}.
For nbits=2, we divide the vectors into sets of 4-dimensional sub-vectors and use 256 centroids for clustering the sub-vectors, while for nbits=1 we set the sub-vector dim to 8 and the same for the rest.
With only 2 bits per dimension, the MRR@10 score on MS MARCO Dev only drops 4\% but the storage is reduced by 83\% and latency is reduced by 76\%.
\begin{table}[!t]
\centering
\begin{adjustbox}{max width=0.48\textwidth}
\begin{tabular}{c|ccc}
\toprule
Condition & MRR@10 & Storage (GB) &Latency (ms)\\\midrule
original &0.362 &78.3 &3.95\\
nbits=2 &0.361 &13.3 &0.90 \\
nbits=1 &0.356 &11.0 &0.92 \\
\bottomrule
\end{tabular}
\end{adjustbox}
\caption{Product Quantization. Pruning threshold is set to 0.9. nbits: ratio of total bytes to the vector dimension.}
\label{tbl:pq}
\end{table}

\begin{figure}[t!]
\centering 
\includegraphics[width=.48\textwidth]{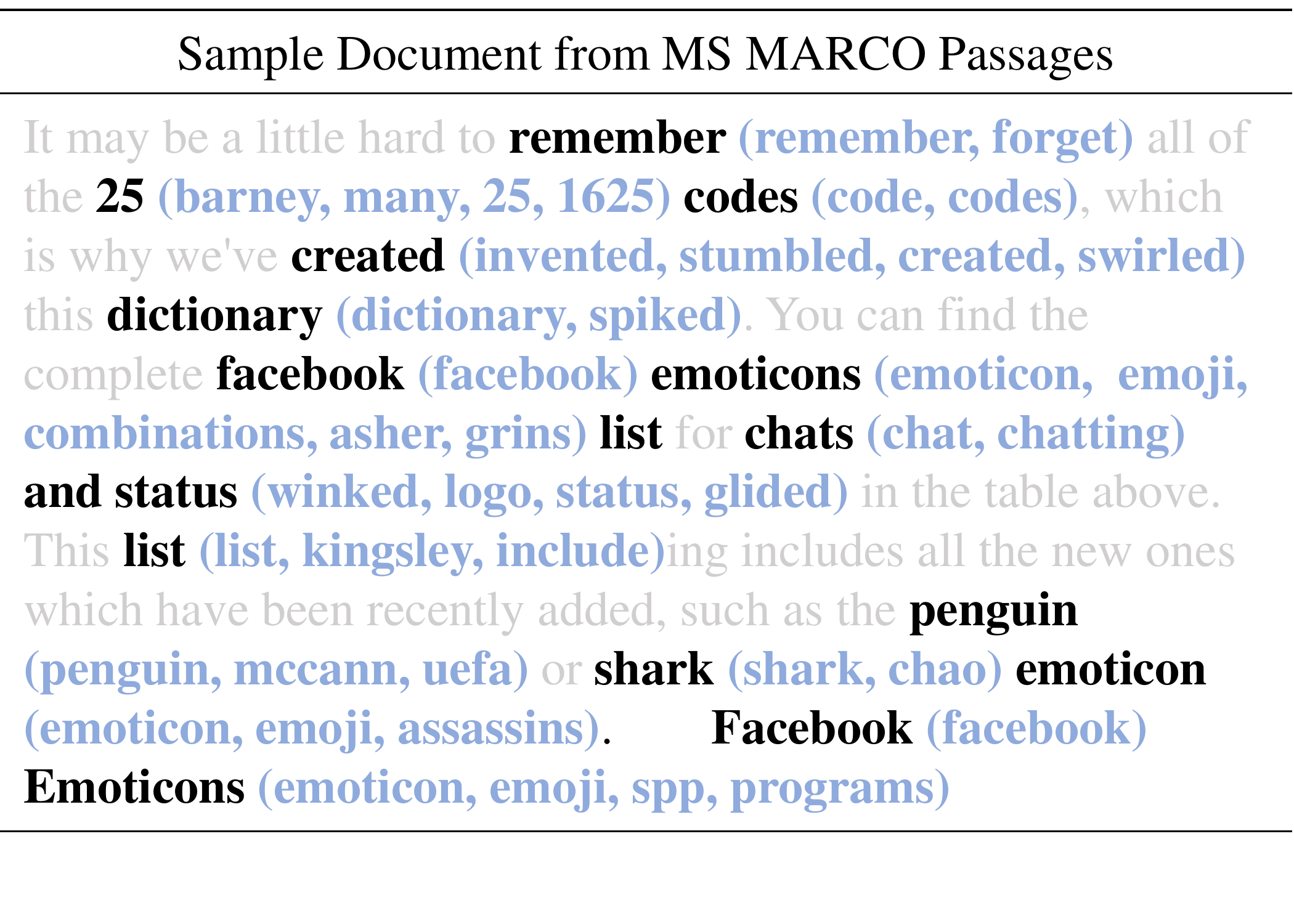}
\vspace{-1cm}
\caption{Attribution analysis of CITADEL. Grey tokens are deactivated, while bold tokens are routed to at least one activated key (in blue).}
\label{fig:facebook_example}
\end{figure}
\subsection{Token Routing Analysis of CITADEL}\label{sec:interpret}
\begin{figure*}[t!]
\centering
\includegraphics[width=1\textwidth]{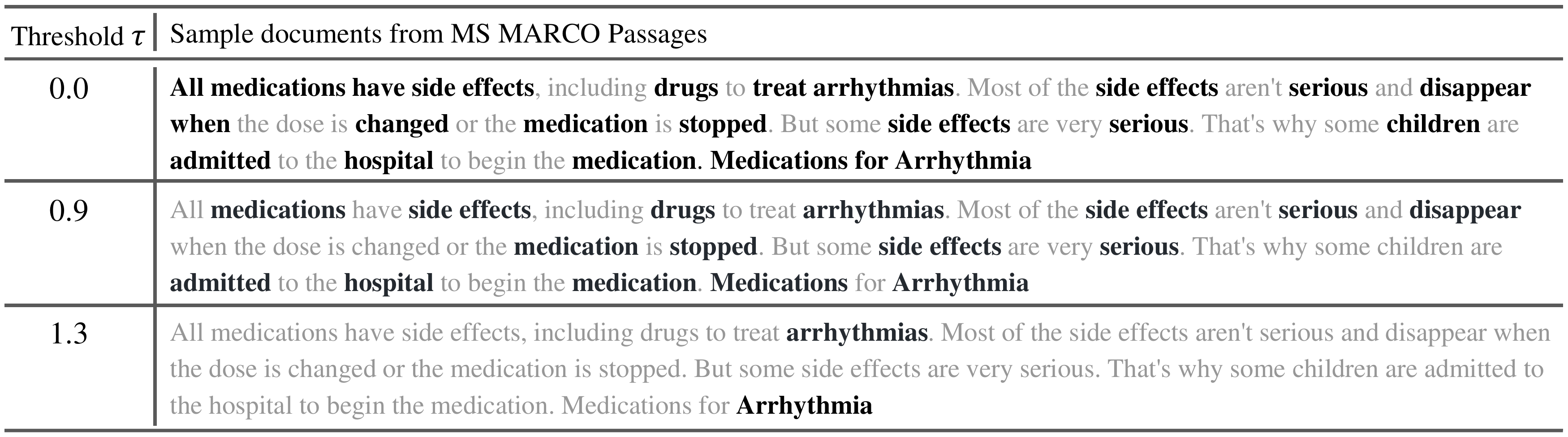}
\caption{Tokens in grey (pruned) have zero activated keys while bold tokens have at least one activated key. We leave out the expanded terms and routing weights due to space limit.}
\label{fig:arrythmia}
\end{figure*}

\paragraph{Qualitative Analysis.} We further analyze the token representations of CITADEL to better understand its predictions.
Figure~\ref{fig:facebook_example} shows an example from MS MARCO passages using the default pruning threshold $\tau=0.9$.
We can see that a lot of redundant words that do not contribute to the final semantics are deactivated, meaning all their routing weights are 0.
For the activated tokens, we could see the routed keys are contextualized as many of them are related to \emph{emoji} which is the theme of the document.
We could further control the representations with post-hoc pruning as shown in Figure~\ref{fig:arrythmia}.
By increasing the pruning threshold, more keywords are pruned and finally leaving the most central word ``\textit{arrhythmia}'' activated.

\begin{figure}[t!]
\centering
\hspace{-0.7cm}
\includegraphics[width=.51\textwidth]{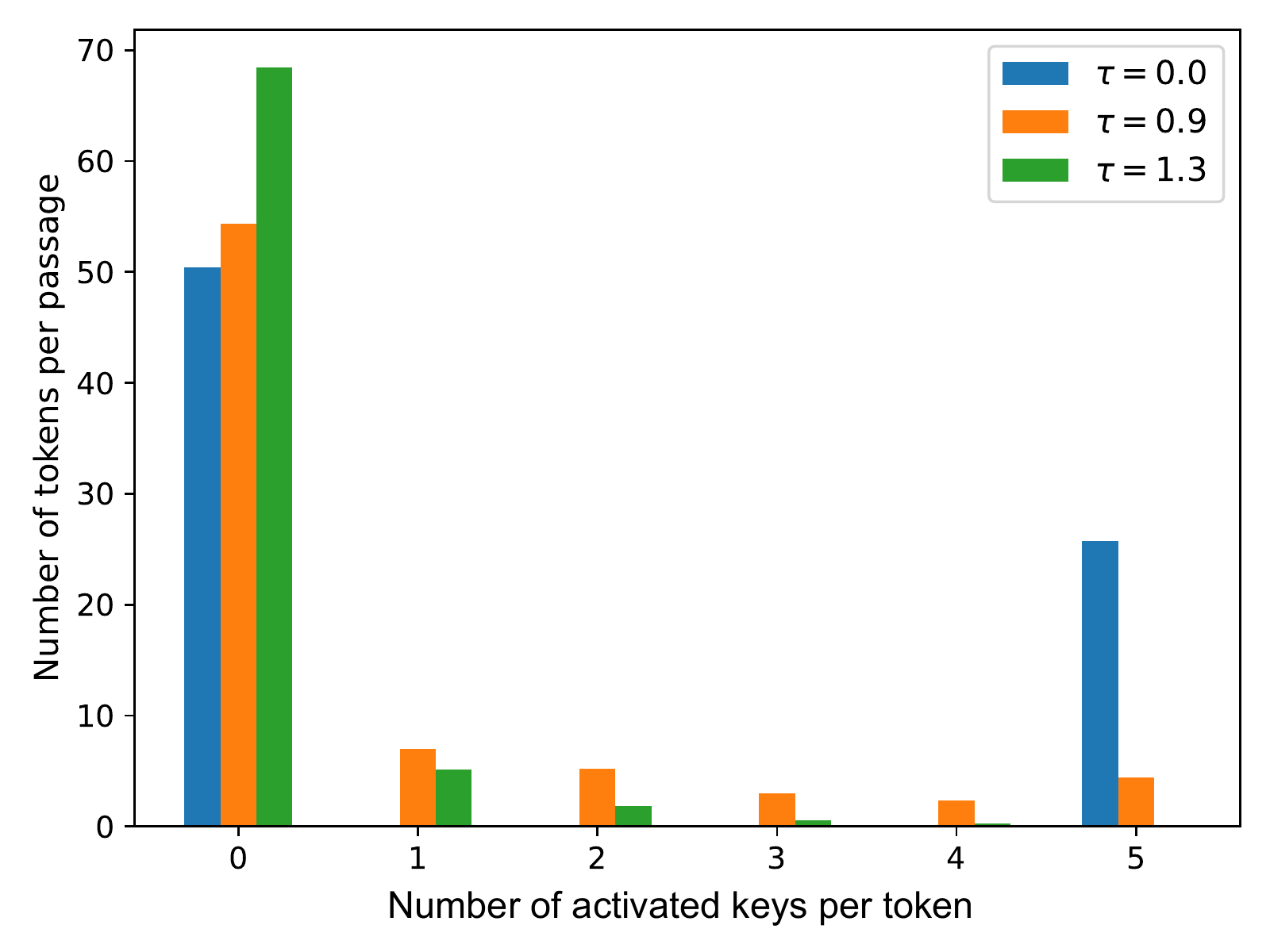}
\caption{Token number distribution over number of activated experts per passage. $\tau$ is the pruning threshold.}
\label{fig:token_dist}
\end{figure}
\paragraph{Quantitative Analysis.} We also perform a quantitative analysis on the token distribution over the number of activated routing keys as shown in Figure~\ref{fig:token_dist}.
We plot the histogram of the number of tokens over the number of activated keys for the whole corpus.
With $\ell_1$ regularization, we already have around 50 tokens per passage in average deactivated, which means that all the routing weights of these 50 tokens are 0.
As the pruning threshold increases, more and more tokens are deactivated, yielding a sparse representation for interpreting CITADEL's behaviours.

\section{Ablation Study}\label{sec:analysis:ablation}
\paragraph{Impact of \texttt{[CLS]}}
Table~\ref{tbl:router_ablation} shows the influence of removing the \texttt{[CLS]} vector for CITADEL on MS MARCO passage.
Although removing \texttt{[CLS]} improves the latency by a large margin, the in-domain effectiveness is also adversely affected, especially on TREC DL 2019.
Nevertheless, CITADEL-tok (w/o \texttt{[CLS]}) still outperforms its counterpart COIL-tok in both precision and latency.

\paragraph{Number of Routed Experts.}
Table~\ref{tbl:num_experts} shows the influence of changing the maximum number of keys that each document token can be routed to during training and inference on MS MARCO passage.
As the number of routing keys increases, the index storage also increases rapidly but so does the MRR@10 score which plateaus after reaching 7 keys.
The latency does not increase as much after reaching 3 routing keys which is probably due to the load balancing loss.

\begin{table}[!t]
\centering
\begin{adjustbox}{max width=0.48\textwidth}
\begin{tabular}{l|ccc}
\toprule
Models &Dev &DL19 & Latency (ms)\\ 
\midrule
COIL-full &0.353 &0.704 &47.9\\
COIL-tok &0.350 &0.660 &46.8\\\hline
CITADEL &0.362 &0.687 &3.95\\
CITADEL-tok &0.360 &0.665 &1.64\\
\bottomrule
\end{tabular}
\end{adjustbox}
\caption{\texttt{[CLS]} ablation on MS MARCO passage. We set the pruning threshold to 0.9. MRR@10 is reported for Dev and nDCG@10 is reported for DL19.}
\label{tbl:router_ablation}
\end{table}
\begin{table}[!t]
\centering
\begin{adjustbox}{max width=0.48\textwidth}
\begin{tabular}{c|ccc}
\toprule
\#Keys & MRR@10 &Storage (GB) &Latency (ms)\\\midrule
1 &0.347 &53.6 &1.28 \\
3 &0.360 &136 &14.7 \\
5 &0.364 &185 &18.6 \\
7 &0.370 &196 &20.4 \\
9 &0.367 &221 &19.6 \\
\bottomrule
\end{tabular}
\end{adjustbox}
\caption{Number of routing keys for documents during training. No post-hoc pruning is applied.}
\vspace{-0.1cm}
\label{tbl:num_experts}
\end{table}

\section{Related Works}\label{sec:related}
\paragraph{Dense Retrieval.} Supported by multiple approximate nearest neighbour search libraries~\citep{faiss,scann}, dense retrieval~\citep{karpukhin-etal-2020-dense} gained much popularity due to its efficiency and flexibility.
To improve effectiveness, techniques such as hard negative mining~\cite{ance2021xiong,star} and knowledge distillation~\cite{lin-etal-2021-batch, tasb} are often deployed. 
Recently, retrieval-oriented pre-training\cite{gao-etal-2021-simcse, lu-etal-2021-less, gao-callan-2021-condenser, contriever, gao-callan-2022-unsupervised} also draws much attention as they could substantially improve the fine-tuning performance of downstream tasks. 

\paragraph{Sparse Retrieval.} Traditional sparse retrieval systems such as BM25~\citep{robertson2009bm25} and tf--idf~\citep{Salton_Buckley_1988} represent the documents as a bag of words with static term weights. 
Recently, many works leverage pre-trained language models to learn contextualized term importance~\cite{sparterm, deepimpact,Formal2021SPLADESL,unicoil}.
These models could utilize existing inverted index libraries such as Pyserini~\citep{lin2021pyserini} to perform efficient sparse retrieval or even hybrid with dense retrieval~\cite{colberter, unifier, dhr}.

\paragraph{Multi-Vector Retrieval.} As mentioned in previous sections, ColBERT~\citep{colbert,santhanam-etal-2022-colbertv2,santhanam2022plaid,colberter} is probably the most renowned family in multi-vector retrieval. 
Tons of engineering and optimization has been made to accelerate the retrieval latency and reduce the index storage such as centroid pruning and fast kernels. COIL~\citep{gao-etal-2021-coil} shows another possibility to accelerate token-level retrieval combined with the exact match prior and inverted vector search.
ME-BERT~\citep{luan-etal-2021-sparse} and MVR~\citep{zhang-etal-2022-multi} propose to use a fixed number of token embeddings for late interaction (e.g., top-k positions or special tokens). 
Concurrently to this work, ALIGNER~\citep{Qian2022MultiVectorRA} proposes to frame multi-vector retrieval as a sparse alignment problem between query tokens and document tokens.
They use entropy-regularized linear programming to find the best alignment scheme to trade index size off against effectiveness.
Our 110M model achieves higher in-domain and out-of-domain accuracy than their base and even large variants.
\section{Conclusion}\label{sec:conclusion}
This paper proposes a novel multi-vector retrieval method that achieves state-of-the-art performance on several benchmark datasets while being 40$\times$ faster than ColBERT-v2 and 17$\times$ faster than the most efficient multi-vector retrieval library to date, PLAID.
By jointly optimizing for the token index size and load balancing, our new dynamic lexical routing scheme greatly reduces the redundancy in all-to-all token interaction of ColBERT while bridging the word-mismatch problem in COIL.
Experiments on both in-domain and out-of-domain datasets demonstrate the effectiveness and efficiency of our model.
\bibliography{anthology,custom}
\bibliographystyle{acl_natbib}

\clearpage
\appendix
\section{Appendix}
\label{sec:appendix}
\subsection{Baselines for Section~\ref{sec:exp}}\label{apx:baselines}
All the baseline models below are trained and evaluated under the same setting of CITADEL (e.g., datasets, hyperparameters, and hardwares).
\paragraph{Sparse Retrievers.} BM25~\citep{robertson2009bm25} uses the term frequency and inverted document frequency as features to compute the similarity between documents.
SPLADE~\citep{Formal2021SPLADESL,Formal2021SPLADEVS} leverages pre-trained language model's MLM layer and ReLU activation to yield sparse term importance.
\paragraph{Dense Retrievers.} DPR~\citep{karpukhin-etal-2020-dense} encodes the input text into a single vector. coCondenser~\citep{gao-callan-2022-unsupervised} pre-trains DPR in an unsupervised fashion before fine-tuning.
\paragraph{Multi-Vector Retrievers.} ColBERT~\citep{colbert,santhanam-etal-2022-colbertv2} encodes each token into dense vectors and performs late interaction between query token vectors and document token vectors. COIL~\citep{gao-etal-2021-coil} applies an exact match constraint on late interaction to improve efficiency and robustness. 

\subsection{Training}\label{apx:training}
For CITADEL, we use bert-base-uncased as the initial checkpoint for fine-tuning.
Following COIL, we set the \texttt{[CLS]} vector dimension to 128, token vector dimension to 32, maximal routing keys to 5 for document and 1 for query, $\alpha$ and $\beta$ in Equation~\eqref{eq:loss} are set to be 1e-2 and 1e-5, respectively.
We add the dot product of \texttt{[CLS]} vectors in Equation~\eqref{eq:dpr_sim} to the final similarity score in Equation~\eqref{eq:citadel_sim}.
All models are trained for 10 epochs with AdamW~\citep{Loshchilov2019DecoupledWD} optimizer, a learning rate of 2e-5 with 3000 warm up steps and linear decay.
Hard negatives are sampled from top-100 BM25 retrieval results.
Each query is paired with 1 positive and 7 hard negatives for faster convergence. 
We use a batch size of 128 on MS MARCO passages with 32 A100 GPUs.

For a fair comparison with recent state of the art models, we further train CITADEL using cross-encoder distillation and hard negative mining.
First, we use the trained CITADEL model under the setting in the last paragraph to retrieve top-100 candidates from the corpus for the training queries.
We then use the cross-encoder\footnote{\url{https://huggingface.co/cross-encoder/ms-marco-MiniLM-L-6-v2}} to rerank the top-100 candidates and score each query-document pair.
Finally, we re-initialize CITADEL with bert-base-uncased using the positives and negatives sample from the top-100 candidates scored by the cross-encoder, with a 1:1 ratio for the soft-label and hard-label loss mixing~\citep{hinton2015distilling}.
We also repeat another round of hard negative mining and distillation but it does not seem to improve the performance any further.

\begin{figure}[t!]
\centering
\vspace{-0.2cm}
\hspace{-0.4cm} 
\includegraphics[width=.5\textwidth]{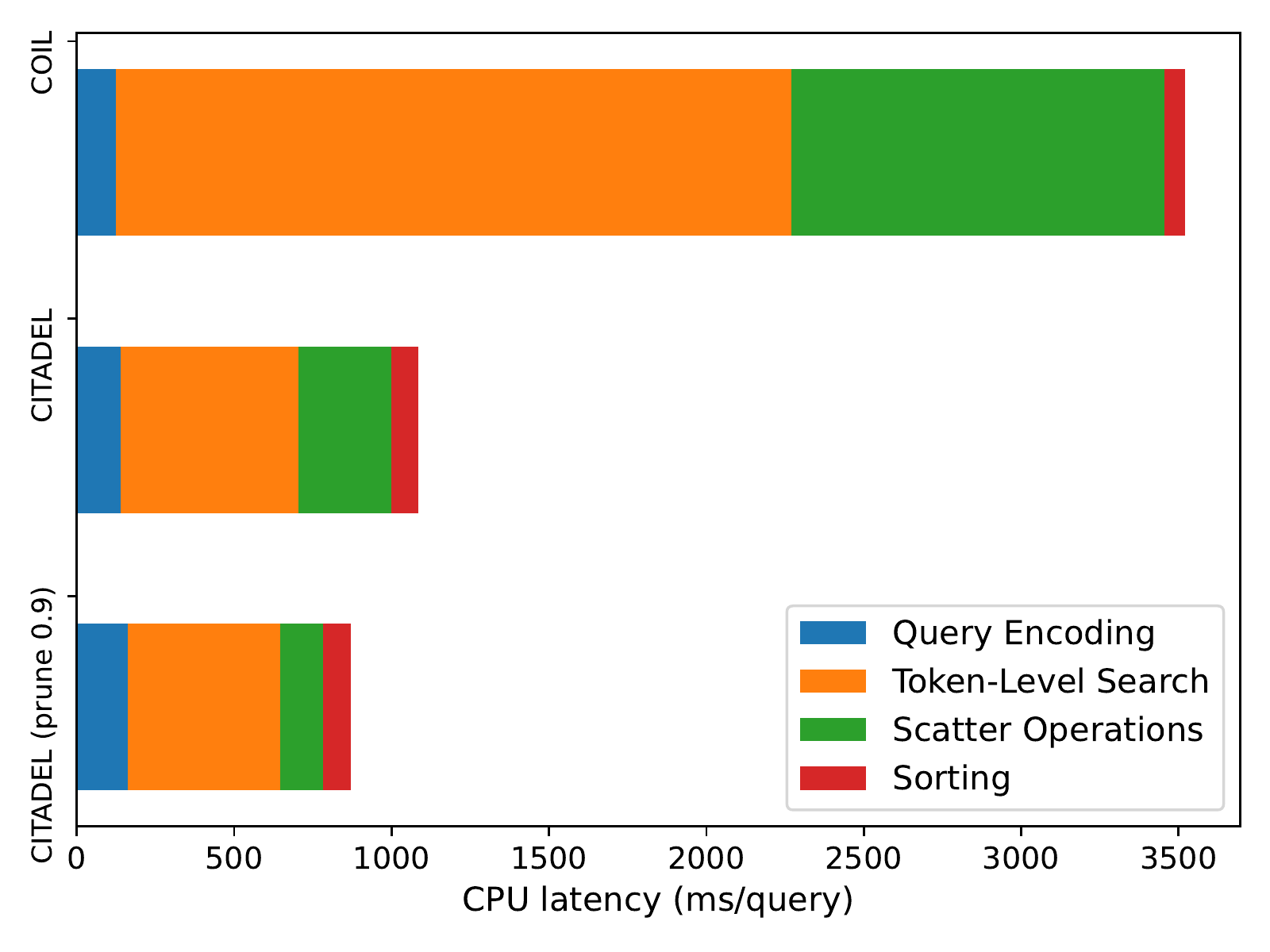}
\caption{Latency breakdown of inverted vector retrieval for CITADEL and COIL. }
\label{fig:latency_breakdown}
\end{figure}
\subsection{Inference and Latency Breakdown}\label{apx:inference}
\paragraph{Pipeline.} We implemented the retrieval pipeline with PyTorch (GPU) and Numpy (CPU), with a small Cython extension module for scatter operations similar to COIL's\footnote{\url{https://github.com/luyug/COIL/tree/main/retriever}}.
As shown in Fig~\ref{fig:latency_breakdown}, our pipeline could be roughly decomposed into four independent parts: query encoding, token-level retrieval, scatter operations, and sorting.
We use the same pipeline for COIL's retrieval process.
For ColBERT's latency breakdown please refer to~\citet{santhanam2022plaid}.
The cost of query encoding comes from the forward pass of the query encoder, which could be independently optimized using quantization or weight pruning for neural networks.
Except that, the most expensive operation is the token-level retrieval, which is directly influenced by the token index size.
We could see that a more balanced index size distribution as shown in Figure~\ref{fig:index_size} has a much smaller token vector retrieval latency.
The scatter operations are used to gather the token vectors from the same passage ids from different token indices, which is also related to the token index size distribution.
Finally, we sort the aggregated ranking results and return the candidates.

\paragraph{Hardwares and Latency Measurement.} 
We measure all the retrieval models in Table~\ref{tbl:msmarco} on a single A100 GPU for GPU search and a single Intel(R) Xeon(R) Platinum 8275CL CPU @ 3.00GHz for CPU search.
All indices are stored in fp32 (token vectors) and int64 (corpus ids if necessary) on disk. 
We use a query batch size of 1 and return the top-1000 candidates by default to simulate streaming queries.
We compute the average latency of all queries on MS MARCO passages' Dev set and then report the minimum average latency across 3 trials following PLAID~\citep{santhanam2022plaid}.
I/O time is excluded from the latency but the time of moving tensors from CPU to GPU during GPU retrieval is included.
\end{document}